\documentclass[useAMS,usenatbib]{mn2e}
\usepackage{graphicx}
\usepackage{rotating}   % used to e.g. rotate figures
\title[Asteroseismology of \mbox{$\theta$~Oph}: constraints on global stellar parameters and core overshooting ]{An asteroseismic study of the $\beta$ Cephei star \mbox{$\theta$~Ophiuchi}:  constraints on global stellar parameters and core overshooting} 

\author[M. Briquet et al.]
{\parbox{179mm}{\begin{flushleft}
\vspace{-0.5cm}
{\LARGE M. Briquet$^1$}\thanks{E-mail:maryline@ster.kuleuven.be}\thanks{Postdoctoral Fellow of the Fund for Scientific Research, Flanders},
{\LARGE T. Morel$^1$}\thanks{European Space Agency (ESA) postdoctoral external fellow},
{\LARGE A. Thoul$^{2}$}\thanks{Chercheur qualifi\'e FNRS},
{\LARGE R. Scuflaire$^{2}$},
{\LARGE A. Miglio$^{2}$},
{\LARGE J. Montalb\'an$^{2}$},
{\LARGE M.-A. Dupret$^{3}$},
{\LARGE and C. Aerts$^{1,4}$}
\end{flushleft}
}\vspace*{0.200cm}\\  
\parbox{159mm}{
$^1$ Instituut voor Sterrenkunde, Katholieke Universiteit Leuven, Celestijnenlaan 200 D, B-3001 Leuven, Belgium\\
$^2$ Institut d'Astrophysique et de G\'eophysique de Li\`ege, Universit\'e de Li\`ege, all\'ee du Six Ao\^ut 17, B-4000 Li\`ege, Belgium\\
$^3$ Observatoire de Paris, LESIA, 5 place Jules Janssen, 92195 Meudon Principal Cedex, France\\
$^4$ Department of Astrophysics, University of Nijmegen, PO Box 9010, 6500 GL Nijmegen, the Netherlands}}

\begin{document}

\date{}

\pagerange{\pageref{firstpage}--\pageref{lastpage}} \pubyear{2007}

\maketitle

\label{firstpage} 

\begin{abstract}
We present a seismic study of the $\beta$~Cephei star $\theta$~Ophiuchi. Our analysis is based on the observation of one radial mode, one rotationally split $\ell = 1$ triplet and three components of a rotationally split $\ell = 2$ quintuplet for which the $m$-values were well identified by spectroscopy. We identify the radial mode as fundamental, the triplet as $p_1$ and the quintuplet as $g_1$. Our NLTE abundance analysis results in a metallicity and CNO abundances in full agreement with the most recent updated solar values. With $X \in [0.71,0.7211]$ and $Z \in [0.009,0.015]$, and using the Asplund et al.\ (\cite{asplund}) mixture but with a Ne abundance about 0.3 dex larger (Cunha et al.\ \cite{cunha}), the matching of the three independent modes, enables us to deduce constrained ranges for the mass ($M$ = 8.2$\pm$0.3 M$_\odot$) and central hydrogen abundance ($X_c$ = 0.38$\pm$0.02) of $\theta$~Oph and to prove the occurrence of core overshooting ($\alpha_{ \rm ov}$ = 0.44$\pm$0.07). We also derive an equatorial rotation velocity of 29$\pm$7 km s$^{-1}$. Moreover, we show that the observed non-equidistance of the $\ell=1$ triplet can be reproduced by second order effects of rotation. Finally, we show that the observed rotational splitting of two modes cannot rule out a rigid rotation model.
\end{abstract}

\begin{keywords}
stars: early-type -- stars: individual: \mbox{$\theta$~Oph} -- stars: oscillations -- stars: abundances -- stars: interiors
\end{keywords}

\section[]{Introduction}
For a few years many breakthroughs have been achieved in the field of asteroseismology of $\beta$~Cephei stars thanks to several observational photometric and spectroscopic multi-site campaigns and long-term monitoring dedicated to this kind of pulsating B-type stars. The aim is to improve our knowledge of their internal structure and more precisely of two mixing processes that affect their evolutionary path: core overshooting and internal rotation. 

The first detailed asteroseismic modelling was performed for \mbox{V836~Cen}, which led to constraints on global stellar parameters but also on the core overshooting parameter. Moreover the non-rigid rotation of the star was proved (Aerts et al.\ \cite{aerts}, Dupret et al.\ \cite{dupret}). Similar results were afterwards obtained for $\nu$ Eri (Pamyatnykh et al.\ \cite{pamyatnykh}, Ausseloos et al.\ \cite{ausseloos}). Recently, Aerts et~al.\ (\cite{aerts3}) gave constraints on the physical parameters of $\delta$~Ceti thanks to the discovery of low-amplitude modes by the satellite MOST. Finally, the seismic interpretation by Mazumdar et al.\ (\cite{mazumdar}) showed the occurrence of core overshooting for $\beta$~CMa. The derived overshooting parameter values are 0.10$\pm$0.05, 0.05$\pm$0.05, 0.20$\pm$0.05 and 0.20$\pm$0.05 for V836 Cen, $\nu$ Eri, $\delta$~Ceti and $\beta$~CMa, respectively.

The $\beta$~Cephei star \mbox{$\theta$~Ophiuchi} was also the subject of intensive photometric and spectroscopic observations as described in Handler et al.\ (\cite{handler}) and Briquet et al.\ (\cite{briquet}) (hereafter Paper\,I and Paper\,II), respectively. It was found that \mbox{$\theta$~Oph} has a frequency spectrum which is similar to that of \mbox{V836~Cen}. In this paper we present our modelling based on accurate frequency determination and successful mode identitification obtained in Paper\,I and II. Our main objective is to test if the occurrence of core overshooting and non-rigid rotation found for \mbox{V836~Cen} also applies to \mbox{$\theta$~Oph}.
   
The paper is organized as follows. In Section\,2 we summarize the observational pulsation constraints which constitute the starting point of our seismic modelling of \mbox{$\theta$~Oph}. In Section\,3 we perform a detailed abundance analysis of \mbox{$\theta$~Oph} with the aim to use the deduced metallicity as an additional constraint. In Section\,4 we present the evolution and oscillation codes that we used in our study, as well as the physical inputs. In Section\,5 we derive the seismic constraints on global stellar parameters and on core overshooting. In Section\,6 we test the hypothesis of a non-rigid rotation model thanks to the two observed multiplets. We end with a conclusion in Section\,7.    

\section[]{Observational constraints}
The observational pulsation characteristics of \mbox{$\theta$~Oph} derived in Paper\,I and II can be summarized as follows. The photometric data (Paper\,I) were gathered in the framework of a three-site campaign, allowing the detection of seven pulsation frequencies. The identification of their corresponding $\ell$-value showed the presence of one radial mode, one rotationally split $\ell = 1$ triplet and three components of a rotationally split $\ell = 2$ quintuplet. In addition, the spectroscopic observations (Paper\,II) lifted the ambiguity for the $m$-value of the observed $\ell = 2$ main mode. The pulsation frequencies and their $(\ell,m)$-values are listed in Table\,\ref{log}. We note that such a frequency spectrum was observed for the star \mbox{V836~Cen} (Aerts et al.\ \citet{aerts2}). We also point out that the mode identifications of all the observed components of the quintuplet are determined for \mbox{$\theta$~Oph}, which was not the case for \mbox{V836~Cen}.

\begin{figure}  
\includegraphics[bb = 60 50 520 410, width=8cm]{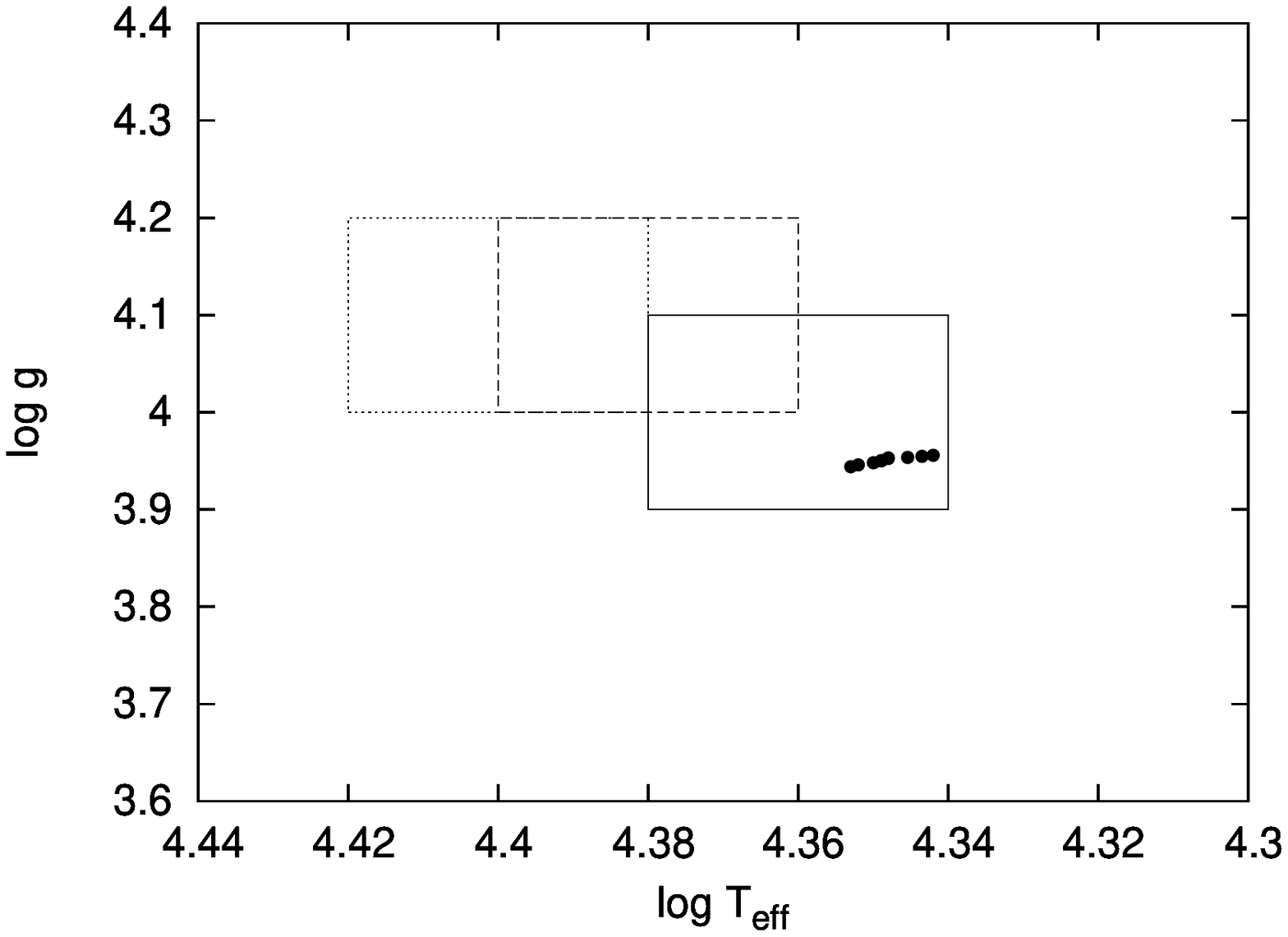}  
\caption{The error boxes represent the position of $\theta$~Oph in the HR diagram as derived from photometric (full line, Paper\,I) and spectroscopic data (dashed line, Paper\,II; dotted line, this paper). The positions of the models which fit exactly the three independent modes are also shown for the different couples ($Z,\alpha_{\rm ov}$) given in Table\,\ref{Z_alpha}. }  
\label{HR}  
\end{figure}

The position of appropriate models will be compared with the position of the star in the HR diagram that was determined photometrically and spectroscopically in Paper\,I and II. The obtained error boxes are represented in Fig.\,\ref{HR}. We note that such a deviation between photometrically and spectroscopically derived effective temperatures is common for B-type stars (e.g.\ De Ridder et al.\ \cite{deridder}, Morel et al.\ \cite{morel}). 

Recently, Niemczura \& Daszynska-Daszkiewicz\ (\cite{niemczura}) determined [M/H] for \mbox{$\theta$~Oph}. However, they did not have information that \mbox{$\theta$~Oph} is a triple system composed of a B2 primary, a spectroscopic secondary with a mass lower than 1 M$_\odot$ (Briquet et al.\ \cite{briquet}) and a speckle B5 star (McAlister et al.\ \cite{McAlister}). In what follows, we present a careful abundance analysis for the primary by taking into account the presence of the tertiary, the contribution to the lines of the secondary being negligible.

\begin{table}
\caption{The pulsation modes derived from the photometric and spectroscopic results presented in Paper\,I and Paper\,II, respectively. The amplitudes of the modes are given for the $u$ filter and for the radial velocities.}
\begin{center}
\begin{tabular}{cccccc}
\hline\hline
ID & Frequency (d$^{-1})$ & $(\ell,m)$ & $u$ ampl. & RV ampl. \\
 & & &   (mmag) &  (km s$^{-1}$)\\
\hline
$\nu_1$ & 7.1160 & $(2,-1)$ &  12.7 &  2.54 \\
$\nu_2$ & 7.2881 & $(2,1)$ & 2.1 & $-$ \\
$\nu_3$ & 7.3697 & $(2,2)$ &  3.6 & $-$ \\
$\nu_4$ & 7.4677 & $(0,0)$ &  4.7 &   2.08\\
$\nu_5$ & 7.7659 & $(1,-1)$ &  3.4 & $-$ \\
$\nu_6$ & 7.8742 & $(1,0)$ & 2.3 &  $-$ \\
$\nu_7$ & 7.9734 & $(1,1)$ &  2.4 & $-$  \\
\hline
\end{tabular}
\end{center}
\label{log}
\end{table}
 
\section{Abundance analysis}\label{sect_abundances}
The non-local thermodynamic equilibrium (NLTE) abundances of He, C, N, O, Mg, Al, Si, S and Fe were calculated using the latest versions of the line formation codes DETAIL/SURFACE and plane-parallel, fully line-blanketed Kurucz atmospheric models (Kurucz\ \cite{kurucz93}). Curve-of-growth techniques were used to determine the abundances using the equivalent widths of a set of unblended lines measured in a mean CORALIE spectrum (see Paper\,II), which was created by co-adding the 86 individual exposures (all put in the laboratory rest frame prior to this operation). The reader is referred to Morel et al.\ (\cite{morel}) for complete details on the methodology used to derive the elemental abundances. 

To correct for the contamination of the spectrum by the tertiary, our study is based on synthetic, composite spectra assuming the following parameters for this companion: $T_{\rm eff}$=19\,000 K, $\log g$=4.0 dex [cgs] (Paper\,II) and a microturbulent velocity, $\xi$=5 km s$^{-1}$, typical of B-type dwarfs. We also considered that the tertiary contributes to 22\% of the total light of the system in the optical band (Paper\,I). In addition, we assumed the star to have a chemical composition typical of OB dwarfs in the solar vicinity (Daflon \& Cunha\ \cite{daflon_cunha}). For iron, we assumed an abundance $\log \epsilon$(Fe)=7.3 dex (Morel et al.\ \cite{morel}). We will examine below the sensitivity of our results to these assumptions. 

A standard, iterative scheme is first used to self-consistently derive the atmospheric parameters: $T_{\rm eff}$ is determined from the Si\,II/III ionization balance, $\log g$ from fitting the collisionally-broadened wings of the Balmer lines and $\xi$ from requiring the abundances yielded by the O\,II features to be independent of the line strength. We obtain: $T_{\rm eff}$=25\,000$\pm$1000 K, $\log g$=4.10$\pm$0.15 dex [cgs] and $\xi$=4$_{-3}^{+2}$ km s$^{-1}$. For comparison, we obtained in Paper\,II, $T_{\rm eff}$=24\,000$\pm$1000 K and $\log g$=4.1$\pm$0.1 dex [cgs] using the NLTE code FASTWIND (Puls et al.\ \cite{puls}). Other studies show that using different methods on the same dataset can indeed lead to uncertainties of order 500\,K for B-type stars (e.g.\ Smalley \& Dworetsky \cite{smalley}, Morel et al.\ \cite{morel}, Kaiser \cite{kaiser}). 

The abundances are given in Table~\ref{tab_abundances} and are compared with the standard solar mixture of Grevesse \& Sauval\ (\cite{grevesse_sauval}) and with values derived from time-dependent, three-dimensional hydrodynamical models (Asplund et al.\ \cite{asplund}, and references therein). The quoted uncertainties take into account both the line-to-line scatter and the errors arising from the uncertainties on the atmospheric parameters. Note that a possible downward revision of $T_{\rm eff}$ by $\sim$1000 K (see above) is explicitly taken into account in the total error budget. We infer a low helium content, but this quantity is uncertain and may be considered solar within the large error bars. On the other hand, there is no indication for the nitrogen excess occasionally observed in other slowly-rotating $\beta$ Cephei stars (Morel et al.\ \cite{morel}). The resulting metallicity, $Z$=0.0114$\pm$0.0028, is identical, within the errors, to the most recent and likely realistic estimates for the Sun (Table~\ref{tab_abundances}). To calculate this quantity, the abundances of the elements not under study were taken from Grevesse \& Sauval\ (\cite{grevesse_sauval}). Our assumed neon abundance is indistinguishable from recent values derived for a sample of B-type stars in the Orion association (Cunha et al.\ \cite{cunha}). The other species are trace elements and contribute only negligibly to metal mass fraction. 

\begin{table}
\centering
\caption{Mean NLTE abundances (on the scale in which $\log \epsilon$[H]=12) and total 1-$\sigma$ uncertainties. The number of used spectral lines is given in brackets. For comparison purposes, we provide in the last two columns the standard solar composition of Grevesse \& Sauval\ (1998; Sun 1-D) and updated values in the present day solar photosphere derived from three-dimensional hydrodynamical models (Asplund et al.\ 2005; Sun 3-D).}

\begin{tabular}{lccc} \hline\hline
 & $\theta$ Oph & Sun 1-D & Sun 3-D\\\hline
He/H                & 0.066$\pm$0.026 (10) & 0.085$\pm$0.001   & 0.085$\pm$0.002\\
$\log \epsilon$(C)  & 8.32$\pm$0.09 (7)    & 8.52$\pm$0.06     & 8.39$\pm$0.05\\ 
$\log \epsilon$(N)  & 7.78$\pm$0.10 (23)   & 7.92$\pm$0.06     & 7.78$\pm$0.06\\ 
$\log \epsilon$(O)  & 8.58$\pm$0.26 (27)   & 8.83$\pm$0.06     & 8.66$\pm$0.05\\ 
$\log \epsilon$(Mg) & 7.49$\pm$0.15  (2)   & 7.58$\pm$0.05     & 7.53$\pm$0.09\\ 
$\log \epsilon$(Al) & 6.24$\pm$0.14  (4)   & 6.47$\pm$0.07     & 6.37$\pm$0.06\\ 
$\log \epsilon$(Si) & 7.04$\pm$0.22  (8)   & 7.55$\pm$0.05     & 7.51$\pm$0.04\\ 
$\log \epsilon$(S)  & 7.22$\pm$0.27  (5)   & 7.33$\pm$0.11     & 7.14$\pm$0.05\\ 
$\log \epsilon$(Fe) & 7.41$\pm$0.17 (27)   & 7.50$\pm$0.05     & 7.45$\pm$0.05\\
$Z$                 & 0.0114$\pm$0.0028    & 0.0172$\pm$0.0012 & 0.0124$\pm$0.0007\\
\hline
\end{tabular}
\label{tab_abundances}
\end{table}

To examine the sensitivity of our results to the various assumptions made about the physical properties of the companion, we have repeated the abundance analysis after varying the adopted effective temperature, surface gravity, chemical composition, and luminosity of the tertiary within the range of plausible values. Namely, we assumed in turn: $T_{\rm eff}$=21\,000 K, $\log g$=3.7 dex [cgs], the abundances of all the metals enhanced by 0.3 dex relative to solar and a contribution of only 18\% to the total light of the system in the optical, while keeping the other parameters unchanged. As expected, the abundances of the chemical elements determined from lines of low-ionization ionic species (e.g. Mg, S) are most strongly affected by the choice of the parameters for the cool component (Table~\ref{tab_sensitivity}). However, the metallicity remains largely unaltered in all cases. Our conclusions regarding the metal content of $\theta$ Oph appear therefore robust against the exact nature of its speckle companion. 

\begin{table}
\centering
\caption{Sensitivity of the derived metal abundances and metallicity of $\theta$ Oph on the assumed properties of the tertiary. We quote the abundance differences compared with the values listed in Table~\ref{tab_abundances}.}
\begin{tabular}{lcccc} \hline\hline
                          & $\Delta$$T_{\rm eff}$= & $\Delta$$\log g$= & $\Delta\log \epsilon$= & flux ratio= \\
                          & +2000 K                & --0.3 dex         & +0.3 dex               & 18\%\\\hline
$\Delta\log \epsilon$(C)  & --0.05   & --0.02    & --0.05   & --0.01\\ 
$\Delta\log \epsilon$(N)  & --0.05   & --0.02    & --0.03   & --0.02\\ 
$\Delta\log \epsilon$(O)  & --0.06   & --0.02    & --0.03   & --0.03\\ 
$\Delta\log \epsilon$(Mg) &  +0.06   &  +0.02    & --0.15   &  +0.02\\ 
$\Delta\log \epsilon$(Al) & --0.05   & --0.01    & --0.04   & --0.01\\ 
$\Delta\log \epsilon$(Si) &  +0.05   &  +0.01    & --0.15   &  +0.03\\ 
$\Delta\log \epsilon$(S)  &  +0.12   &  +0.00    & --0.44   &  +0.05\\ 
$\Delta\log \epsilon$(Fe) & --0.04   & --0.02    & --0.04   & --0.02\\ 
$\Delta$Z                 & --0.0007 & --0.0003  & --0.0011 & --0.0003\\
\hline
\end{tabular}
\label{tab_sensitivity}
\end{table}
  
\section[]{Stellar models}
The numerical tools and physical inputs used in our study are the following. The stellar models for non-rotating stars were computed with the evolutionary code CL\'ES (Code Li\'egeois d'\'Evolution Stellaire, Scuflaire et al.\ \cite{scuflaire1}). We used the OPAL2001 equation of state (Rogers \& Nayfonov\ \cite{rogers}) and Caughlan \& Fowler\ (\cite{caughlan}) nuclear reaction rates with Formicola et al.\ (\cite{formicola}) for the $^{14}$N$(p,\gamma)^{15}$\,O cross-section. Convective transport is treated by using the classical Mixing Length Theory of convection (B\"ohm-Vitense\ \cite{bohm}). As shown in the previous section, the abundances of $\theta$~Oph are in full agreement with the solar values of Asplund et al.\ (\cite{asplund}). For the chemical composition, we consequently used the solar mixture from these authors, except for Ne. For this latter element, a direct abundance determination in a small sample of nearby B stars using photospheric lines (Cunha et al.\ \cite{cunha}) suggests a value $\sim$0.3 dex larger than quoted by Asplund et al.\ (\cite{asplund}). For our computations, we consequently adopted $\log \epsilon$(Ne)=8.11. We used OP opacity tables (Seaton\ \cite{seaton}) computed for the mixture in Cunha et al.\ (\cite{cunha}) (that is the mixture of Asplund et al.\ (\cite{asplund}) and $\log \epsilon$(Ne)=8.11). These tables are completed at $\log T < 4.1$ with the low temperature tables of Ferguson et al.\ (\cite{ferguson}) for the Asplund et al.\ (\cite{asplund}) mixture, as the effect of increasing Ne on low temperature opacities can be neglected for such a hot star. We computed stellar models with and without taking into account microscopic diffusion. For models with diffusion, we used the formulation of Thoul et al.\ (\cite{thoul}).

Stellar models are parametrized by the initial hydrogen abundance $X$, the core convective overshooting parameter $\alpha_{\rm ov}$, the metallicity $Z$, the mass $M$ and the central hydrogen abundance $X_c$ (which is related to the age).

For each stellar model, we calculated the theoretical frequency spectrum of low-order $p$- and $g$-modes with a degree of the oscillation up to $\ell = 2$ using a standard adiabatic code (Scuflaire et al.\ \cite{scuflaire2}), which is much faster than a non-adiabatic code but leads to the same theoretical pulsation frequencies within the adopted precision of the fit, which was 10$^{-3}$ d$^{-1}$. Once the models fitting the observed modes are selected, we checked the excitation of the pulsation modes with the linear non-adiabatic code MAD developed by Dupret et al.\,(\cite{dupret0}).

In an attempt to explain the asymmetries of the observed multiplets, we also computed the adiabatic frequencies with the code FILOU (Tran Minh \& L\'eon\ \cite{tran}), which includes effects of rotation up to the second order, according to the formalism of Soufi et al.\ (\cite{soufi}).

\section[]{Constraints on stellar parameters and core overshooting}
\subsection{Effects of diffusion}
$\theta$~Oph is a slow rotator with an equatorial rotation velocity of about 30 km s$^{-1}$ (Paper\,II). Moreover, its surface convection zone is very thin. In such conditions, diffusion mechanisms can occur and alter the photospheric abundances. We consequently investigated if diffusion could be the explanation for the marginal lower He content of $\theta$~Oph compared to the solar value. Moreover, we checked its effect on the oscillation frequencies. 

\begin{figure}  
\includegraphics[bb = 0  150 600 740, width=8.7cm]{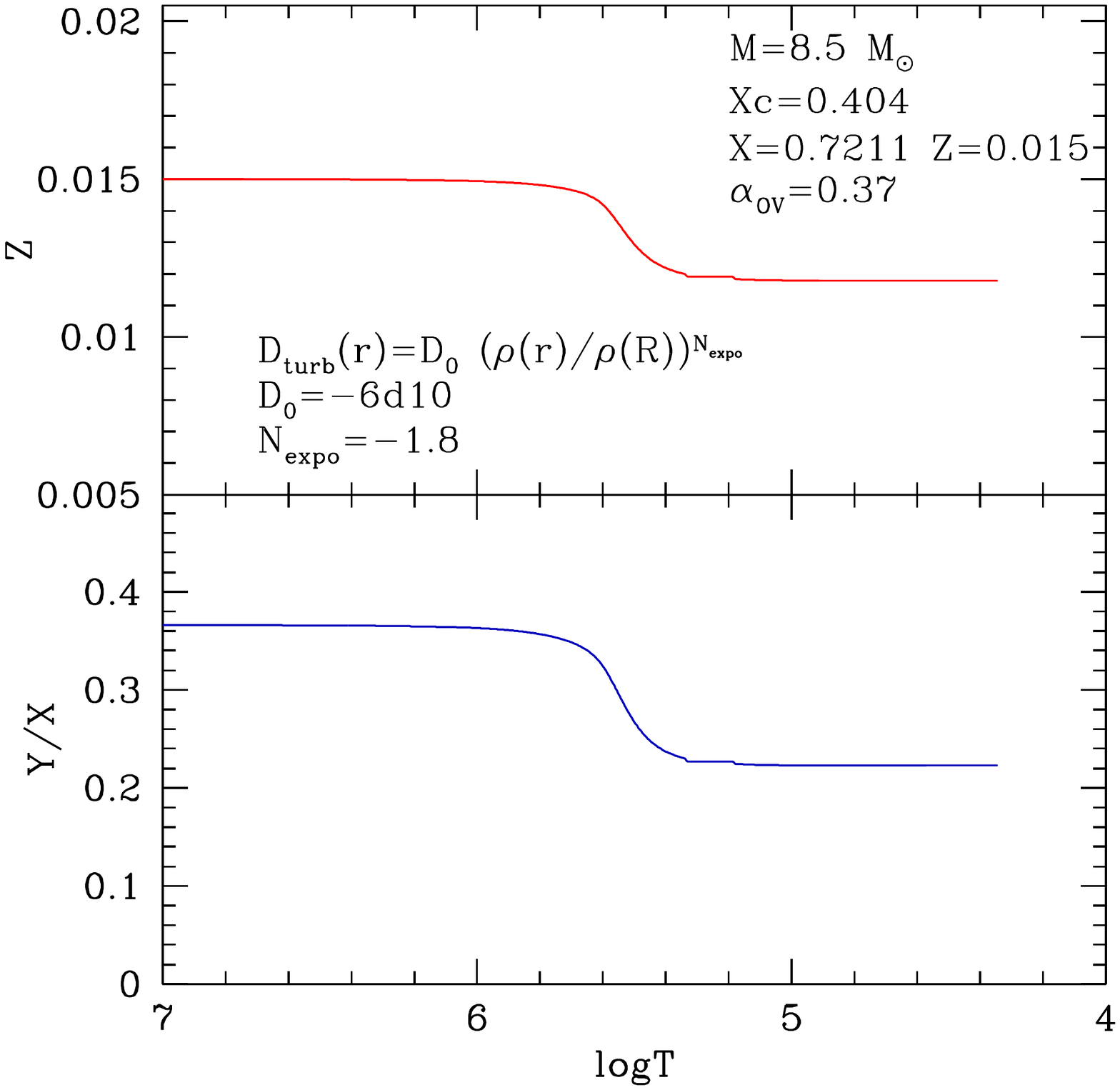}  
\caption{Metallicity and Y/X profiles for a model which fits the three observed frequencies and has both a solar initial composition and a surface composition compatible with the observations.}  
\label{diff}  
\end{figure}

Our calculations include microscopic diffusion (without radiative forces and wind, and using TBL94's routine; Thoul et al.\ \cite{thoul}) and a turbulent mixing consistent with the results of Talon et al.\ (\cite{talon}). We were able to reproduce the observed surface metallicity and helium abundances with models having initially the solar composition ($X=0.7211$, $Y=0.264$, $Z=0.01485$). Those models are very close to those obtained without diffusion, because the diffusion only affects the very superficial layers of the star. In particular, the models calculated with and without diffusion have exactly the same frequency spectrum. In Fig.\,\ref{diff} we show the metallicity and the $Y/X$ profiles for a model which fits the three observed frequencies and has both a solar initial composition and a surface composition compatible with the observations. The diffusion and the turbulent mixing only affect layers down to a radius of 0.92 $R$.

Even though these results agree very well with the observations, it might be hazardous to trust them blindly. Indeed, radiative accelerations and stellar winds are very important in those stars and can strongly affect the surface abundances (Bourge et al.\ \cite{bourge}). They were ignored here because calculations involving those effects are computationally intensive and difficult to perform, and clearly beyond the scope of this paper.
 
\subsection{Seismic analysis}

Since taking into account diffusion or not does not affect the derived stellar parameters of our models, we continued our analysis without diffusion but considering sufficiently large ranges for $X$ and $Z$.

For our seismic analysis we first searched for models that fit the radial mode with frequency $\nu_4$ together with the zonal $\ell = 1$ mode with frequency $\nu_6$. We then made use of the quintuplet to add additional constraints.  

\begin{figure}  
\includegraphics[bb = 50 50 580 430, width=8.7cm]{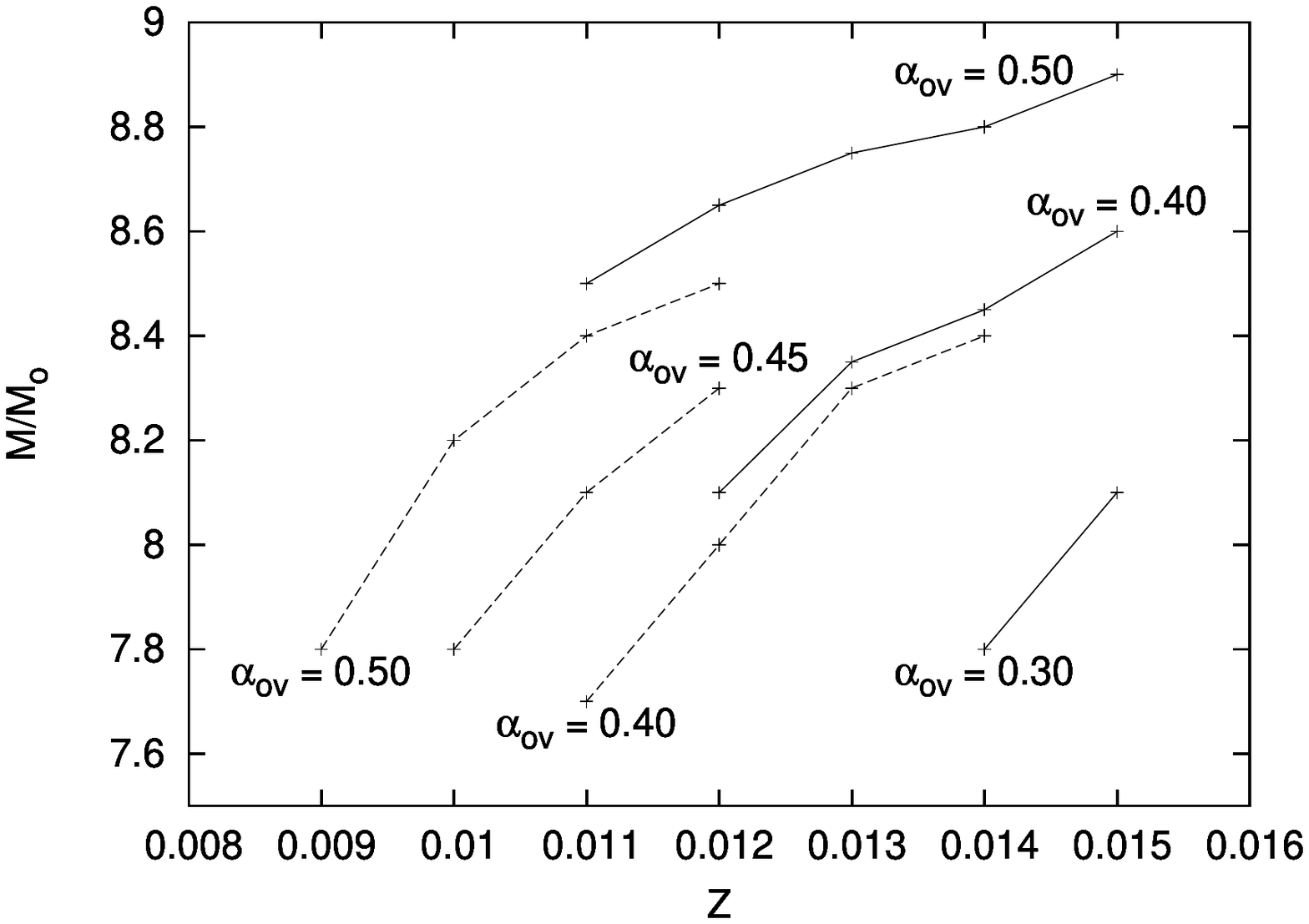}  
\caption{The $M-Z$ relations obtained by matching the radial mode and the central peak of the triplet for several values of the core overshooting parameter, $X$ being fixed to 0.71 (dashed lines) or 0.7211 (full lines).}  
\label{M_Z}  
\end{figure}  

\begin{table}
\caption{Relation between $Z$ and $\alpha_{\rm ov}$ imposed by the matching of the three independent modes.}
\begin{center}
\begin{tabular}{cc}
 \hline\hline
$Z$ & $\alpha_{\rm ov}$ \\
\hline\\[-5pt]      
0.009 & 0.51\\
0.010 & 0.48\\
0.011 & 0.45\\
0.012 & 0.43 \\
0.013 & 0.40 \\
0.014 & 0.38\\
0.015 & 0.37\\
\hline
\end{tabular}
\end{center}
\label{Z_alpha}
\end{table}

\begin{table}
\caption{Physical parameters of the model that matches the observed modes, $X \in [0.71,0.7211]$ and $Z \in [0.009,0.015]$.}
\begin{center}
\begin{tabular}{ccc}
\hline\hline
$M$ ($M_\odot$) & $=$ & 8.2$\pm$0.3\\
$T_{\rm eff}$ (K)& $=$ &  22260$\pm$280\\
$\log g$ (dex)& $=$ & 3.950$\pm$0.006\\
$X_c$ & $=$ & 0.38$\pm$0.02\\
$\alpha_{\rm ov}$ & $=$ & 0.44$\pm$0.07\\
\hline
\end{tabular}
\end{center}
\label{par}
\end{table}

We found that the radial mode is either the fundamental mode or the first overtone. However, the models with $\nu_4$ as the first overtone are further away from the observational position in the HR diagram than the models with $\nu_4$ as fundamental. Moreover, none of the three modes is excited by the classical $\kappa$ mechanism for models with $\nu_4$ as the first overtone, even for a value of $Z$ of 0.015 with Y of 0.0264. We consequently concluded that the radial mode is identified as fundamental and a scan of stellar parameter space also reveals that the triplet is identified as $p_1$.

Fitting one frequency results in finding one model along the evolutionary track for every combination of $(X,\ \alpha_{\rm ov},\ Z,\ M)$. Then, $X$ being fixed, fitting two frequencies gives a relation between two parameters for given values of the last one. For the considered values of $\alpha_{\rm ov}$ one thus gets several $M-Z$ relations that are shown in Fig.\,\ref{M_Z}. We can see that the fitting of the two frequencies implies an increase in mass if either the metallicity or the core convective overshooting parameter increases. This figure also illustrates the order of magnitude difference in mass induced by different adopted $X$ values.

Because the zonal mode of the quintuplet was not observed, we computed it as the average frequency of the two surrounding modes with frequency $\nu_1$ and $\nu_2$. A scan of stellar parameter space shows that the quintuplet is identified as $g_1$. The fitting of a third independent frequency implies a relation between the metallicity and the core overshooting parameter: the lower the metallicity the higher the overshooting. This $\alpha_{\rm ov}-Z$ relation is presented in Table\,\ref{Z_alpha}. 

The positions in the HR diagram of the models reproducing the observed modes are shown in Fig.\,\ref{HR} for several couples ($Z,\alpha_{\rm ov}$). One can see that the derived models are situated in the cooler part of the photometric observed error box, and outside the spectroscopic one. This is actually the case for all studied $\beta$~Cephei stars up to now and needs to be futher investigated.
By considering a wide range of metallicities $Z \in [0.009,0.015]$, for $X \in [0.71,0.7211]$, one obtains a core overshooting parameter $\alpha_{\rm ov} \in [0.37,0.51]$ and a mass $M \in [7.9,8.5]$ $M_\odot$. The other physical parameters are given in Table\,\ref{par}. The three modes are well excited by the classical $\kappa$ mechanism for a metallicity larger than 0.011. However, radiatives forces on iron allow this element to accumulate in the excitation region and lead to the excitation of additional modes in lower metallicity stars (Pamyatnykh et al.\ \cite{pamyatnykh}, Bourge et al.\ \cite{bourge}). Note that the only other theoretically excited mode with $\ell \le 2$ is the $f$-mode with $\ell = 1$ but only for $Z$ larger than 0.013.

We point out that models matching the observed modes but computed using the solar abundances of Grevesse \& Sauval\,(\cite{grevesse_sauval}) with OPAL opacity tables are not excited for $Z \sim 0.01$. We refer to Miglio, Montalb\'an \& Dupret\,(\cite{miglio}) for a detailed discussion on the implication of the adopted opacity tables and metal mixtures on the excitation of pulsation modes.

We also mention that the amount of overshooting found for $\theta$~Oph corresponds exactly to that computed by Deupree (\cite{deupree}) by means of 2D hydrodynamic simulations of zero-age main-sequence convective cores. Our derived value is also in agreement with results obtained by Ribas et al. (\cite{ribas}) who provided an empirical calibration of convective core overshooting for a range of stellar masses by studying eight detached double-lined eclipsing binaries. They found a systematic increase of the amount of convective overshooting with the stellar mass, the values being 0.3-0.6 for $\sim$10 $M_\odot$ stars. 

\section{Constraints on the rotation}

\subsection{First order analysis}

When the rotation frequency is small compared to both $\sqrt{R^3/GM}$ and the
considered pulsation frequency, the pulsation frequencies $\nu_m$
of modes differing only by the $m$ value of the spherical functions are linked
through a simple relation. If we assume that the rotational frequency
$\nu_{rot}$ is a function of the radius $r$ only, this relation reads
\begin{equation}
\nu_m=\nu_0+m\int_0^1K(x)\nu_{rot}(x)\,dx\,,
\label{eq_split}
\end{equation}
where $x=r/R$. The rotational kernel $K(x)$ depends on the considered mode
(see Lynden-Bell \& Ostriker\,\cite{lynden_bell} or Unno et al.\,\cite{unno}).

The $\ell=1$, $p_1$ triplet and three components (corresponding to $m=-1$, 1 and
2) of the $\ell=2$, $g_1$ quintuplet are observed. The components of these
multiplets are not strictly equidistant as required by equation~(\ref{eq_split}).
These departures from equidistance may result from the fact that the rotation
velocity of the star is too large for a linear approximation to be valid. It may
also result from the fact that a magnetic field contributes to the splitting. As
the departures from equidistance are not too large, we tentatively interpret
them as errors in the measure of the splitting. So, we have
$\Delta\nu_1=$ 0.10375$\pm$0.005 d$^{-1}$ for mode $\ell=1$, $p_1$ and
$\Delta\nu_2=$ 0.08457$\pm$0.003 d$^{-1}$ for mode $\ell=2$, $g_1$, denoting by
$\Delta\nu$ the integral term in equation~(\ref{eq_split}).

\begin{figure}
%\begin{center}
\includegraphics[bb = 40 50 380 310,width=7.9cm]{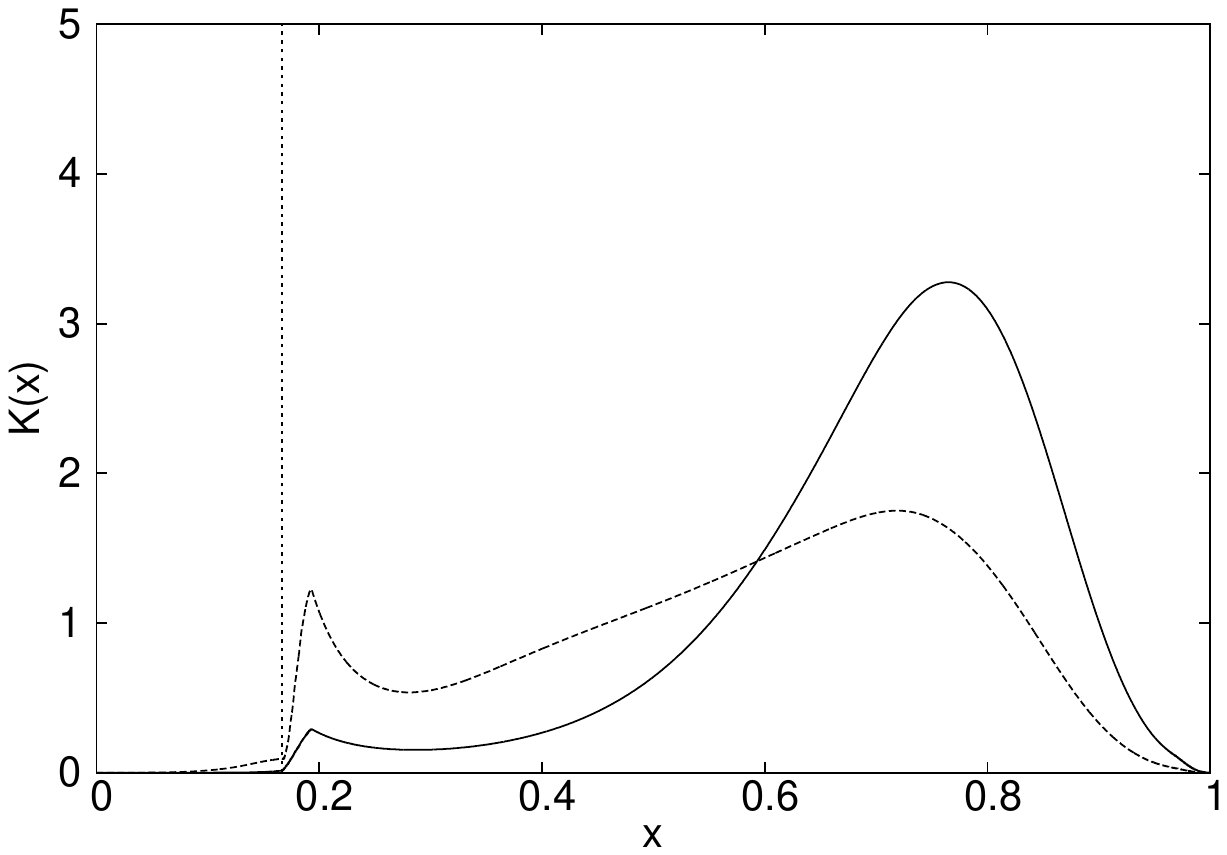}
%\end{center}
\caption{The kernels for the $\ell=1$, $p_1$ mode (solid line) and for the
$\ell=2$, $g_1$ mode (dashed line). The vertical dotted line marks the position of
the boundary of the convective core.}
\label{fig_kern}
\end{figure}

Fig.\,\ref{fig_kern} shows the behaviour of the rotational kernels for the
modes we are investigating. It is clear that they do not probe the convective
core. This was already the case for V836 Cen (Dupret et al.\,\cite{dupret}) and, as in
this case, with just two pieces of information on the behaviour of the rotation
velocity, we analyse its trend inside the envelope by fitting the linear
expression
\begin{equation}
\nu_{rot}(x)=\nu_{rot,0}+\nu_{rot,1}(x-1)\,.
\label{eq_difrot}
\end{equation}
The splittings are then given by
\begin{equation}
\Delta\nu_k=c_{k0}\nu_{rot,0}+c_{k1}\nu_{rot,1}\quad k=1,2
\label{eq_splitsys}
\end{equation}
with $k$ referring to the two known splittings, and
\begin{eqnarray}
c_{k0}&=&\int_0^1K_k(x)\,dx\,,\\
c_{k1}&=&\int_0^1K_k(x)(x-1)\,dx\,.
\end{eqnarray}\\
Taking the errors on the splittings into account, the system~(\ref{eq_splitsys})
gives (all frequencies in d$^{-1}$)\\
\begin{tabular}{ccc}
$0.0878$&$<\nu_{rot,0}$&$<0.1419$\,,\\
$-0.0491$&$<\nu_{rot,1}$&$<0.0971$\,.
\end{tabular}\\
The splitting data are thus consistent with a constant rotation velocity inside
the envelope, a rotation period of 9.2$\pm$2.2 days and an equatorial rotation velocity of 29$\pm$7 km s$^{-1}$. This latter value is in full agreement with the $v\sin i$ and equatorial rotation velocity derived in Paper\,II.

\subsection{Second order analysis}
 
\begin{table}
\caption{Observed rotational splittings (d$^{-1}$). The subscripts of $\nu_{\ell,m}$ denote
the degree $\ell$ and azimuthal order $m$.}
\begin{center}
\begin{tabular}{cccc}
\hline\hline
$\ell=1$ & $\nu_{1,0}-\nu_{1,-1}$ & $\nu_{1,1}-\nu_{1,0}$ & $\nu_{1,1}-\nu_{1,-1}$\\
         & 0.1083 & 0.0992 & 0.2075 \\
$\ell=2$ & $(\nu_{2,1}-\nu_{2,-1})/2$ & $\nu_{2,2}-\nu_{2,1}$ & $\nu_{2,1}-\nu_{2,-1}$\\
         & 0.08605 & 0.0816 & 0.1721  \\
\hline
\end{tabular}
\end{center}
\label{rotobs}
\end{table}

\begin{table}
\caption{Theoretical rotational splittings (d$^{-1}$) obtained with a second order 
perturbative treatment of rotation, for a model fitting the observed zonal modes.}
%rigid rotation with $\nu_{\rm rot}=0.10751\,{\rm d}^{-1}$.}
\begin{center}
\begin{tabular}{cccc}
\hline\hline
\multicolumn{4}{l}{Solid rotation: $\nu_{\rm rot}=0.10751\,{\rm d}^{-1}$} \\
$~~\ell=1$ & $\nu_{1,0}-\nu_{1,-1}$ & $\nu_{1,1}-\nu_{1,0}$ & $\nu_{1,1}-\nu_{1,-1}$\\
         & 0.10934 &  0.09816 & 0.2075 \\
$~~\ell=2$ & $(\nu_{2,1}-\nu_{2,-1})/2$ & $\nu_{2,2}-\nu_{2,1}$ & $\nu_{2,1}-\nu_{2,-1}$\\
         & 0.08661 & 0.08859 & 0.1732  \\[0.2cm]
\multicolumn{4}{l}{Solid rotation: $\nu_{\rm rot}=0.10682\,{\rm d}^{-1}$} \\
$~~\ell=1$ & $\nu_{1,0}-\nu_{1,-1}$ & $\nu_{1,1}-\nu_{1,0}$ & $\nu_{1,1}-\nu_{1,-1}$\\
         & 0.10860 &  0.09756 & 0.2062 \\
$~~\ell=2$ & $(\nu_{2,1}-\nu_{2,-1})/2$ & $\nu_{2,2}-\nu_{2,1}$ & $\nu_{2,1}-\nu_{2,-1}$\\
         & 0.08605 & 0.08800 & 0.1721  \\[0.2cm]
\multicolumn{4}{l}{Differential rotation: $\nu_{\rm rot}=0.10915 \,+\,0.00549\,(x-1)\,{\rm d}^{-1}$} \\
$~~\ell=1$ & $\nu_{1,0}-\nu_{1,-1}$ & $\nu_{1,1}-\nu_{1,0}$ & $\nu_{1,1}-\nu_{1,-1}$\\
         & 0.10932 &  0.09819 & 0.2075 \\
$~~\ell=2$ & $(\nu_{2,1}-\nu_{2,-1})/2$ & $\nu_{2,2}-\nu_{2,1}$ & $\nu_{2,1}-\nu_{2,-1}$\\
         & 0.08605 & 0.08826 & 0.1721\\
\hline
\end{tabular}
\end{center}
\label{rotth}
\end{table}

A small asymmetry is observed in the $\ell=1$ and $\ell=2$ multiplets, 
as shown in Table~\ref{rotobs}. It is well known that asymmetries are explained
by the effect of terms of higher order in $\nu_{\rm rot}/\nu_{\rm puls}$
in the pulsation equations.
We determine here the adiabatic frequencies with the code FILOU 
(Tran Minh \& L\'eon \cite{tran}, Su\'arez \cite{suarez}). In the version of the code used here, the
effects of rotation are included up to the second order, following Soufi et al. ({\cite{soufi}). 
This code needs as input the spherically symmetric component of the structure model; it determines a posteriori the second order deformation due to rotation.
In principle, the gravity must be corrected for the effect of centrifugal acceleration
already in the spherically symmetric component of the model. This correction would
affect very slightly the frequencies (slow rotation) and has a negligible effect on 
the multiplet asymmetries. It is not included here,
which allows us to use as input one of our best no-rotation model fitting the zonal mode
frequencies, as determined in the previous sections. The main global parameters of this model are:
$M=8.4\:M_{\odot}$, $\log(L/L_\odot)=3.7346$,
$T_{\rm eff}=22053$\,K, $X=0.72$, $Z=0.014$ and $\alpha_{\rm ov}=0.38$.

In Table~\ref{rotth}, we give the theoretical rotational splittings
obtained with this model and the second order treatment of rotation.
We recall that the second order terms cancel in the combination $\nu_{\ell,m}-\nu_{\ell,-m}$.
The last column gives such combination, which we use in the fitting procedure.
Comparing columns 1 and 2 shows the splitting asymmetry. 
For the first results given in this table, we consider a rigid rotation. 
In the first case, the rotation frequency is 
$\nu_{\rm rot}=0.10751\,{\rm d}^{-1}$.
With this value, we fit exactly the observed value of the $\ell=1$ splitting:
$\nu_{1,1}-\nu_{1,-1} = 0.2075$ d$^{-1}$.
In the second case, the rotation frequency is 
$\nu_{\rm rot}=0.10682\,{\rm d}^{-1}$.
With this value,  we fit exactly the observed value of the $\ell=2$ splitting:
$\nu_{2,1}-\nu_{2,-1} = 0.1721$ d$^{-1}$.
In the last case of Table~\ref{rotth}, we consider a
differential rotation law of the same linear form as Eq.~(\ref{eq_difrot}).
The coefficients are adjusted to fit at the same time the observed $\ell=1$ and $\ell=2$ splittings
$\nu_{1,1}-\nu_{1,-1}$ and $\nu_{2,1}-\nu_{2,-1}$.
This gives the linear differential rotation law: 
$\nu_{\rm rot}=0.10915 \,+\,0.00549\,(x-1)\,{\rm d}^{-1}$.
As for the first order analysis, we see that rigid rotation models cannot be eliminated.

The results of Table~\ref{rotth} show that the non-equidistance of the $\ell=1$ triplet
is relatively well reproduced by the second order effect of rotation. However,
the asymmetries of the $\ell=2$ multiplet do not fit the observations. This discrepancy 
could come from observations (multiplet not entirely resolved) or theory 
(effect of higher order terms). We did the same analysis with other models fitting the
zonal mode frequencies and find quasi-identical results for the rotation velocity.}

%\multicolumn{5}{c}{Observations}\\

%\begin{table}
%\caption{Observed and theoretical rotational splittings (d$^{-1}$) obtained
%with a second order perturbative treatment.}
%\begin{center}
%\begin{tabular}{cccc}
%\hline\hline
%$\ell=1$ & $\nu_{m=0}-\nu_{m=-1}$ & $\nu_{m=1}-\nu_{m=0}$ & $\nu_{m=1}-\nu_{m=-1}$\\
%         & 0.1083 & 0.0992 & 0.2075 \\
%$\ell=2$ & $(\nu_{m=1}-\nu_{m=-1})/2$ & $\nu_{m=2}-\nu_{m=-1}$ & $\nu_{m=1}-\nu_{m=-1}$\\
%         & 0.08605 & 0.0816 & 0.1721  \\
%\hline
%\end{tabular}
%\end{center}
%\label{rotobs}
%\end{table}

%\begin{table}
%\caption{Theoretical rotational splittings (d$^{-1}$) obtained with a second order 
%perturbative treatment of rotation, for a model fitting the observed zonal modes.}
%\begin{center}
%\begin{tabular}{cccc}
%\hline\hline
%$\ell=1$ & $\nu_{1,0}-\nu_{1,-1}$ & $\nu_{1,1}-\nu_{1,0}$ & $\nu_{1,1}-\nu_{1,-1}$\\
%         & 0.10934 &  0.09816 & 0.2075 \\
%$\ell=2$ & $(\nu_{2,1}-\nu_{2,-1})/2$ & $\nu_{2,2}-\nu_{2,1}$ & $\nu_{2,1}-\nu_{2,-1}$\\
%         & 0.08463 & 0.08595 & 0.08727 & 0.08859 & 0.1732  \\
%\hline
%\end{tabular}
%\end{center}
%\label{rotobs}
%\end{table}

\section[]{Conclusions}
Our study of the $\beta$~Cephei star $\theta$~Ophiuchi is a new illustration of the power of asteroseismology for this class of pulsators. The couple $(X,Z)$ being chosen, the observation of three independent modes is enough to derive the other parameters that characterize a stellar model, for adopted physical inputs. We point out that such a success can also be attributed to the unique derivation of ($\ell,m$) thanks to state-of-the-art empirical mode identification techniques used in Paper\,I and II.

A detailed NLTE abundance analysis showed that the considered abundance values and thus the metallicity of $\theta$~Oph correspond to the new solar mixture of Asplund et~al.\ (\cite{asplund}). In particular, the CNO abundances are much more consistent with the 3D values of Asplund et al.\ (\cite{asplund}) than with the 1D values of Grevesse \& Sauval\ (\cite{grevesse_sauval}). This is generally the case for B-type stars (Morel et al.\ \cite{morel}). We found a mass $M =$ 8.2$\pm$0.3 $M_\odot$ and a central hydrogen abundance $X_c = $ 0.38$\pm$0.02 for the star. $\theta$~Oph is the fifth $\beta$~Cephei star for which the occurrence of core overshooting is deduced by seismic interpretation and is the target with the highest derived value ($\alpha_{\rm ov}$ = 0.44$\pm$0.07) among them. However, it might be that the core overshooting parameter of previously modelled $\beta$~Cephei stars is underestimated, as the case of \mbox{V836~Cen} illustrates it. For the modelling of this star, Dupret et al.\ (\cite{dupret}) adopted a value of $Z$ larger than 0.016 in order to get the excitation of the modes. However, Morel et al.\ (\cite{morel}) determined $Z = 0.0105$$\pm$0.0022 for \mbox{V836~Cen}. In addition, Miglio, Montalb\'an \& Dupret\,(\cite{miglio}) recently showed that modes can be excited for $Z \sim 0.01$ if one uses the new solar abundances together with the OP opacities, which was not the case in Dupret et al.\ (\cite{dupret}). Finally, considering a lower value of $Z$ increases the $\alpha_{\rm ov}$ of the star (see Fig.\,3 in Dupret et al.\ \cite{dupret}).

We also showed that the asymmetry observed in the $\ell = 1$ triplet can be well reproduced by taking into account the effects of rotation up to the second order. For the quintuplet, the agreement is however not as good. Contrary to \mbox{V836~Cen} (Dupret et~al.\ \cite{dupret}) and $\nu$~Eri (Pamyatnykh et~al.\ \cite{pamyatnykh}) for which non-rigid rotation was proven, the observed rotational splitting of two modes for $\theta$~Oph are still compatible with a rigid rotation model. In the near future, we can expect stronger constraints on the internal rotation of $\beta$~Cephei stars from data collected from space missions (e.g. MOST, COROT). With the observation of rotational splitting of many modes having different probing kernels, we aim to determine the internal rotation law of massive B-type stars. 

\section*{Acknowledgments}
We thank MJ Goupil and JC Su\'arez for allowing us to use the code FILOU. T. M. acknowledges financial support from the European Space Agency through a Postdoctoral Research Fellow grant and from the Research Council of Leuven University through grant GOA/2003/04. We also thank an anonymous referee for constructive comments which helped us to significantly improve our paper.

\label{lastpage}

\end{document}